\documentclass[%
 reprint,
superscriptaddress,
 amsmath,amssymb,
aps,
]{revtex4-2}

\usepackage{graphicx}
\usepackage{dcolumn}
\usepackage{bm}
\usepackage[dvipsnames]{xcolor}
\usepackage{mathtools}
\usepackage{tablefootnote}

\newcommand\Tstrut{\rule{0pt}{2.6ex}}         

\begin{document}

\preprint{APS/123-QED}

\title{Frequency stability of $2.5\times10^{-17}$ in a Si cavity with AlGaAs crystalline mirrors}

\author{Dahyeon~Lee}
\email{dahyeon.lee@colorado.edu}
\affiliation{JILA, National Institute of Standards and Technology and the University of Colorado, Boulder, Colorado 80309-0440, USA \\
and Department of Physics, University of Colorado, Boulder, Colorado 80309-0390, USA}

\author{Zoey~Z.~Hu}
\affiliation{JILA, National Institute of Standards and Technology and the University of Colorado, Boulder, Colorado 80309-0440, USA \\
and Department of Physics, University of Colorado, Boulder, Colorado 80309-0390, USA}

\author{Ben~Lewis}
\affiliation{JILA, National Institute of Standards and Technology and the University of Colorado, Boulder, Colorado 80309-0440, USA \\
and Department of Physics, University of Colorado, Boulder, Colorado 80309-0390, USA}

\author{Alexander~Aeppli}
\affiliation{JILA, National Institute of Standards and Technology and the University of Colorado, Boulder, Colorado 80309-0440, USA \\
and Department of Physics, University of Colorado, Boulder, Colorado 80309-0390, USA}

\author{Kyungtae~Kim}
\affiliation{JILA, National Institute of Standards and Technology and the University of Colorado, Boulder, Colorado 80309-0440, USA \\
and Department of Physics, University of Colorado, Boulder, Colorado 80309-0390, USA}

\author{Zhibin~Yao}
\affiliation{JILA, National Institute of Standards and Technology and the University of Colorado, Boulder, Colorado 80309-0440, USA \\
and Department of Physics, University of Colorado, Boulder, Colorado 80309-0390, USA}

\author{Thomas~Legero}
\affiliation{Physikalisch-Technische Bundesanstalt, Bundesallee 100, 38116 Braunschweig, Germany}

\author{Daniele~Nicolodi}
\affiliation{Physikalisch-Technische Bundesanstalt, Bundesallee 100, 38116 Braunschweig, Germany}

\author{Fritz~Riehle}
\affiliation{Physikalisch-Technische Bundesanstalt, Bundesallee 100, 38116 Braunschweig, Germany}

\author{Uwe~Sterr}
\affiliation{Physikalisch-Technische Bundesanstalt, Bundesallee 100, 38116 Braunschweig, Germany}

\author{Jun~Ye}
\email{ye@jila.colorado.edu}
\affiliation{JILA, National Institute of Standards and Technology and the University of Colorado, Boulder, Colorado 80309-0440, USA \\
and Department of Physics, University of Colorado, Boulder, Colorado 80309-0390, USA}

\date{\today}

\begin{abstract}
Developments in ultrastable lasers have fueled remarkable advances in optical frequency metrology and quantum science. A key ingredient in further improving laser frequency stability is the use of low-noise mirror materials such as AlGaAs crystalline coatings. However, excess noise observed with these coatings limits the performance of cryogenic silicon cavities with AlGaAs mirrors to similar levels achieved with conventional dielectric coatings. With a new pair of crystalline coated mirrors in a 6-cm-long cryogenic silicon cavity operated at 17~K, we demonstrate a clear advantage of crystalline coatings over dielectric coatings. The achieved fractional frequency stability of $2.5 \times 10^{-17}$ at 10~s is four times better than expected for dielectric mirrors and corresponds to more than tenfold reduction in the coating mechanical loss factor. We also combine two silicon cavities to demonstrate optical frequency averaging for enhanced stability. In addition, we present a long-term frequency drift record of four cryogenic silicon cavities measured over several years. These results open up realistic prospects for cavity-stabilized lasers with $10^{-18}$ fractional stability, as well as an all-optical timescale with continuously operating optical local oscillators.
\end{abstract}

\maketitle



\textbf{\emph{Introduction}.} Ultrastable optical interferometers form the backbone of optical atomic clocks \cite{ludlow2015optical}, tabletop tests of fundamental physics \cite{kennedy2020precision, muller2003modern, wiens2016resonator, wcislo2016experimental}, and gravitational wave detectors \cite{harry2006thermal, adhikari2020cryogenic}. Cryogenic silicon cavities continue to push the state-of-the-art in optical cavity frequency stability, reaching thermal noise-limited fractional frequency stability of $3.5 \times 10^{-17}$ up to thousands of seconds \cite{oelker2019demonstration, matei2017}. Despite this impressive performance, cavity-stabilized laser frequency noise is still the limiting factor in improving optical clock stability. To improve the frequency stability of optical cavities even further, the fundamental Brownian thermal noise of the mirror coatings needs to be mitigated, for example, by going to lower temperatures \cite{robinson2019crystalline, valencia2024cryogenic, he2023ultra, barbarat2024towards, wiens2016resonator}, enlarging the mode area, utilizing novel mirror coating materials that exhibit lower Brownian noise \cite{cole2013tenfold, truong2023mid, dickmann2023experimental, harry2006titania}, or indirectly scaling down the noise contribution by using a longer cavity \cite{hafner20158, alvarez2019optical, parke2025three}. While all of these approaches are actively pursued, mirror coatings based on stacks of crystalline GaAs/AlGaAs have recently attracted significant attention \cite{cole2013tenfold, kedar2023frequency, yu2023excess, herbers2022transportable, valencia2024cryogenic, kraus2025ultra, zhu2024ultrastable, didier2018ultracompact}, with possible applications reaching as far as next-generation gravitational wave detectors \cite{cole2023substrate}. These mirrors have lower thermal noise compared to conventional dielectric mirrors due to their lower mechanical loss factor \cite{penn2019mechanical, numata2004thermal, kessler2011thermal}, and can achieve high finesse critically required for ultrastable optical cavities. These desirable qualities make crystalline mirrors an attractive candidate to replace conventional dielectric mirror coatings for high-performance optical cavities.

However, our previous investigations on cryogenic silicon cavities with crystalline mirrors revealed several novel noise mechanisms that hinder reaching the Brownian thermal noise \cite{kedar2023frequency, yu2023excess}. One of the excess noise sources was found to be associated with the birefringence of the crystalline coatings, in which two birefringently split cavity modes showed anti-correlated noise at a level significantly higher than the expected thermal noise. Although the birefringent noise could be suppressed to a sufficiently low level by averaging the frequencies of the two cavity modes, the thermal noise floor of the crystalline coatings still could not be reached because of yet another source of excess ``global'' noise that appeared in two independent systems at JILA and PTB. Unlike the Brownian noise, whose correlation length is on the order of the $\mu$m-scale coating thickness, the global noise was correlated over the mm-scale mode area and thus could not be lowered by using a larger mode size \cite{yu2023excess}. The origin of the excess noise is not clear and may be related to coating impurities, defects, or variations in the bond strength, while it could also be the thermal noise of the mirror optical contact area and the cavity supports. Due to this excess global noise, the frequency stabilities of cryogenic silicon cavities with crystalline mirrors have been limited to levels comparable to those of similar cavities with conventional dielectric mirrors, leaving the full potential of crystalline mirrors unfulfilled.

In this work with a 17~K cryogenic silicon cavity, we demonstrate for the first time clear superiority of crystalline $\text{Al}_{0.92}\text{Ga}_{0.08}\text{As}$ coatings over conventional dielectric mirrors. The frequency stability of $2.5 \times 10^{-17}$, expressed in modified Allan deviation, is a factor of 4 better than the thermal noise limit of an equivalent cavity with conventional dielectric $\text{Si}\text{O}_2/\text{Ta}_2\text{O}_5$ coatings. In addition, we implement optical frequency averaging of two state-of-the-art silicon cavities to improve frequency stability. Finally, we compare the long-term drift rates of four silicon cavities in PTB and JILA compiled over more than 10 years.

\textbf{\emph{Frequency stability}.} Our 6-cm-long silicon cavity uses two 1~m radius of curvature mirrors made of alternating layers of crystalline GaAs/AlGaAs on a silicon substrate. The $12.0\text{-}\mu\text{m}$-thick crystalline mirror coating comprises 48.5 repeats of quarter-wave high-index GaAs and low-index $\text{Al}_{0.92}\text{Ga}_{0.08}\text{As}$ layers, beginning and ending with the GaAs layer. The cavity finesse is 470~000 at the operating wavelength of 1542~nm, corresponding to a cavity linewidth of 5~kHz. The double-cone shaped cavity is vertically mounted from its midplane to reduce vibration sensitivity \cite{notcutt2005simple}. Vibration noise from the cryostat is mitigated with the split-plate design and an active vibration isolation table as described in Refs. \cite{zhang2017ultrastable, robinson2019crystalline}, so that vibration noise does not degrade cavity performance at Fourier frequencies below 3~Hz except for discrete peaks at harmonics of the 1~Hz cryostat cycle frequency. The cavity is cooled with a closed-cycle cryostat to 17~K, where the coefficient of thermal expansion of silicon is zero \cite{wiens2014silicon}. Thermal isolation from the environment is achieved with multiple layers of shielding, including a radiation shield, an actively controlled outer shield, and a passive inner shield.  A 1542~nm fiber laser is locked to the cavity with the Pound-Drever-Hall technique \cite{drever1983laser, black2001introduction}. Residual amplitude modulation (RAM) is actively suppressed well below the cavity thermal noise limit with the AOM FM-triplet scheme described in Ref. \cite{kedar2024synthetic}, which allows us to operate the system at shot noise-limited signal to noise ratio with just 90~nW of cavity transmission. The birefringent noise of the crystalline mirrors is canceled by simultaneously locking to two orthogonal polarization modes of the cavity \cite{kedar2023frequency, yu2023excess}. Without the dual-polarization lock, the laser fractional frequency stability is severely degraded to low-$10^{-16}$ level. Technical noise budget for the system is reported elsewhere \cite{kedar2023frequency}, with the only difference being the RAM cancellation scheme \cite{kedar2024synthetic}.

The 6-cm cavity is nominally identical to the one used in our previous publications \cite{kedar2023frequency, yu2023excess}, but with noteworthy changes to mirror coatings. The cavity used in Refs. \cite{kedar2023frequency, yu2023excess} was contaminated during a routine maintenance of the cryostat and the crystalline mirrors had to be replaced. The newly installed crystalline mirrors have a lower transmission than the previous mirrors by using a GaAs/AlGaAs stack with 3 more periods, increasing the coating thickness from $11.3~\mu\text{m}$ to $12.0~\mu\text{m}$. As optical loss from scattering and absorption remains the same at approximately 5~ppm per mirror, finesse increases from 290~000 to 470~000. The birefringent mode splitting also changed from 770~kHz to 890~kHz with the new set of mirrors. No modification was made to the composition of the coating material and the growth process.

\begin{figure}
\includegraphics[width=\linewidth]{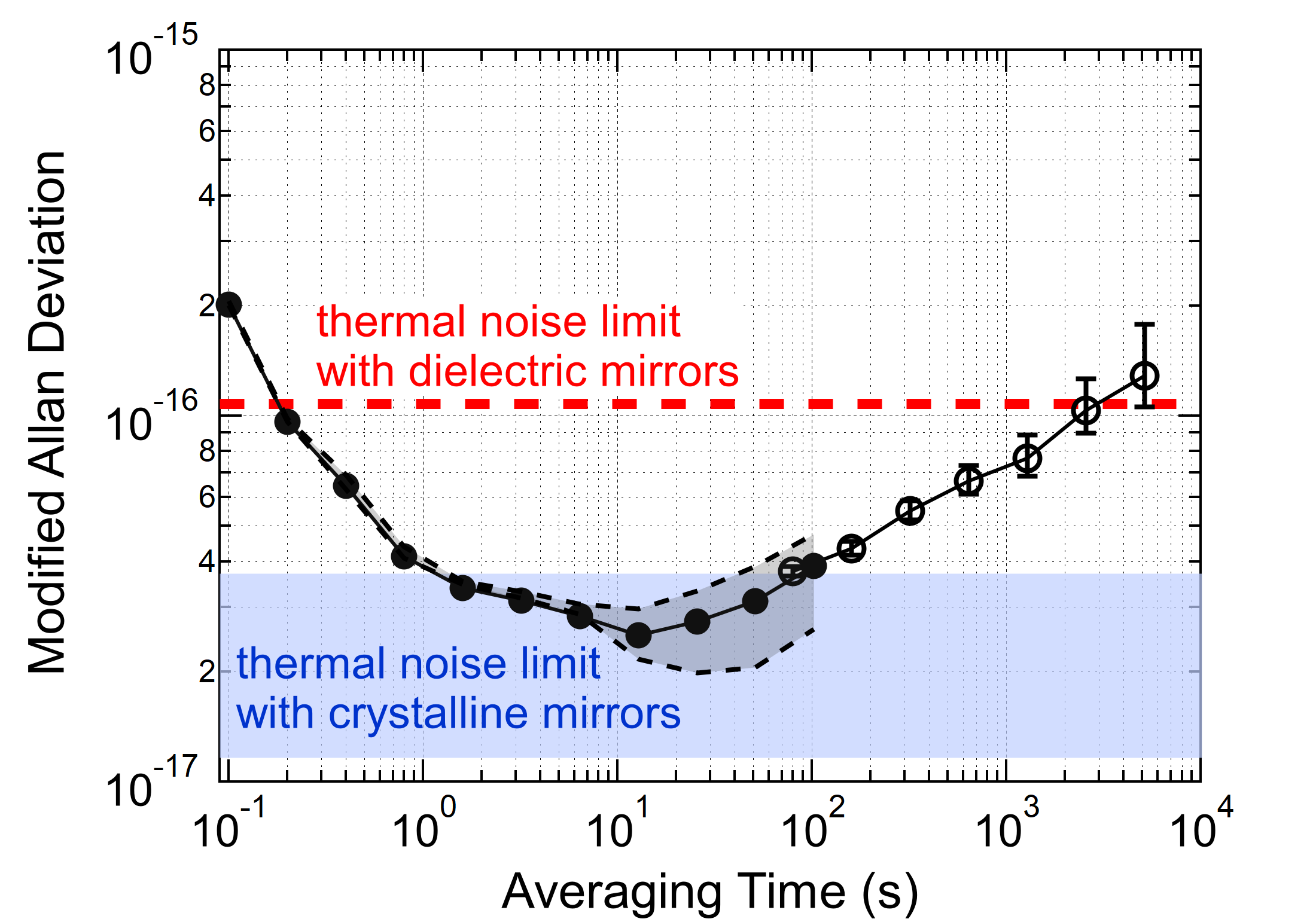}
\caption{\label{fig:mdev}Fractional frequency stability of the 6-cm silicon cavity with crystalline mirrors at 17~K. The use of crystalline mirrors results in a significant reduction of the cavity thermal noise (from red dashed line to blue shaded area, see text). For averaging times less than 100 seconds, the average of 10 three-cornered-hat measurements each lasting 5 hours is shown (filled markers). The gray shaded area shows the full range of these 10 independent measurements. For averaging times longer than 100 seconds, the cavity frequency is measured with a Sr lattice clock (empty markers). Linear drift is removed in all datasets.}
\end{figure}

The fractional frequency stability of the 6-cm cavity is shown in Fig.~\ref{fig:mdev}. Linear drift is removed from all datasets. The frequency stability for averaging times less than 100 seconds (filled markers) is measured with the three-cornered-hat method \cite{gray1974proceedings}, using a 21-cm silicon cavity and a 40-cm ultra-low-expansion (ULE) glass cavity as the other two references \cite{matei2017, swallows2012operating, nicholson2012comparison}. The trace shown is the average of 10 separate three-cornered-hat measurements each lasting 5 hours, and the shaded area marks the full range of observed values. For averaging times longer than 100 seconds, the three-cornered-hat method yields unreliable results due to the relatively high instability of the ULE cavity, so a Sr lattice clock is used for long term measurements (empty markers) \cite{aeppli2024clock}. With the new cavity, the measured frequency stability of $2.5 \times 10^{-17}$ around 10 seconds of averaging is 4 times lower than the expected Brownian thermal noise of conventional dielectric coatings, clearly demonstrating the superior noise performance of crystalline mirrors.

The expected thermal noise level of crystalline mirrors has a large uncertainty because the mechanical loss factor of the coatings is not well characterized, especially at the 17~K operating temperature of our cavity. Published loss factors from mechanical ringdown measurements range from $6 \times 10^{-6}$ to $5 \times 10^{-5}$ in this temperature range \cite{cole2008monocrystalline, cole2012cavity, pag25}, with the correspondingly large uncertainty range of thermal noise displayed as the blue shaded area in Fig.~\ref{fig:mdev}. The loss factor at room temperature is well known to be $2.5 \times 10^{-5}$ \cite{penn2019mechanical}. Our previous work confirmed this value at 124~K by measuring the differential noise between the $\text{HG}_{00}$ and $\text{HG}_{01}$ spatial modes of the same cavity \cite{yu2023excess}. We however identified excess global noise nearly identical for two independent silicon cavities operated at 124~K and 17~K \cite{yu2023excess, kedar2023frequency}, limiting our capability to reduce the thermal noise uncertainty at 17~K. The excess global noise measured in two experiments, after scaling the fractional frequency noise by the respective cavity length, is nearly the same. This leads us to attribute this noise source to the mirror coatings. The latest crystalline coating provided a new opportunity and the cavity performance demonstrated here allows us to put an upper bound on the mechanical loss factor of the crystalline coatings. The observed modified Allan deviation of $\text{2.5}\times10^{-17}$ around 10 s places an upper bound on the 17~K loss of crystalline coatings at $\text{2.3}\times10^{-5}$, which is more than a factor of 10 lower than the 17~K loss of conventional dielectric coatings of $3.2\times10^{-4}$ \cite{robinson2021thermal}.

\textbf{\emph{Optical frequency averaging}.} With two state-of-the-art cryogenic silicon cavities online at JILA, we improved the laser stability even further by averaging the two cavities. While the idea of combining two or more oscillators to achieve better performance is not new \cite{yan2018multi, loh2023optical, sch20d, yan2025high}, it is especially valuable when applied to state-of-the-art silicon cavities because building a new cavity with improved performance is not trivial when the cavity is already at the record performance level.

\begin{figure*}
\includegraphics[width=0.8\linewidth]{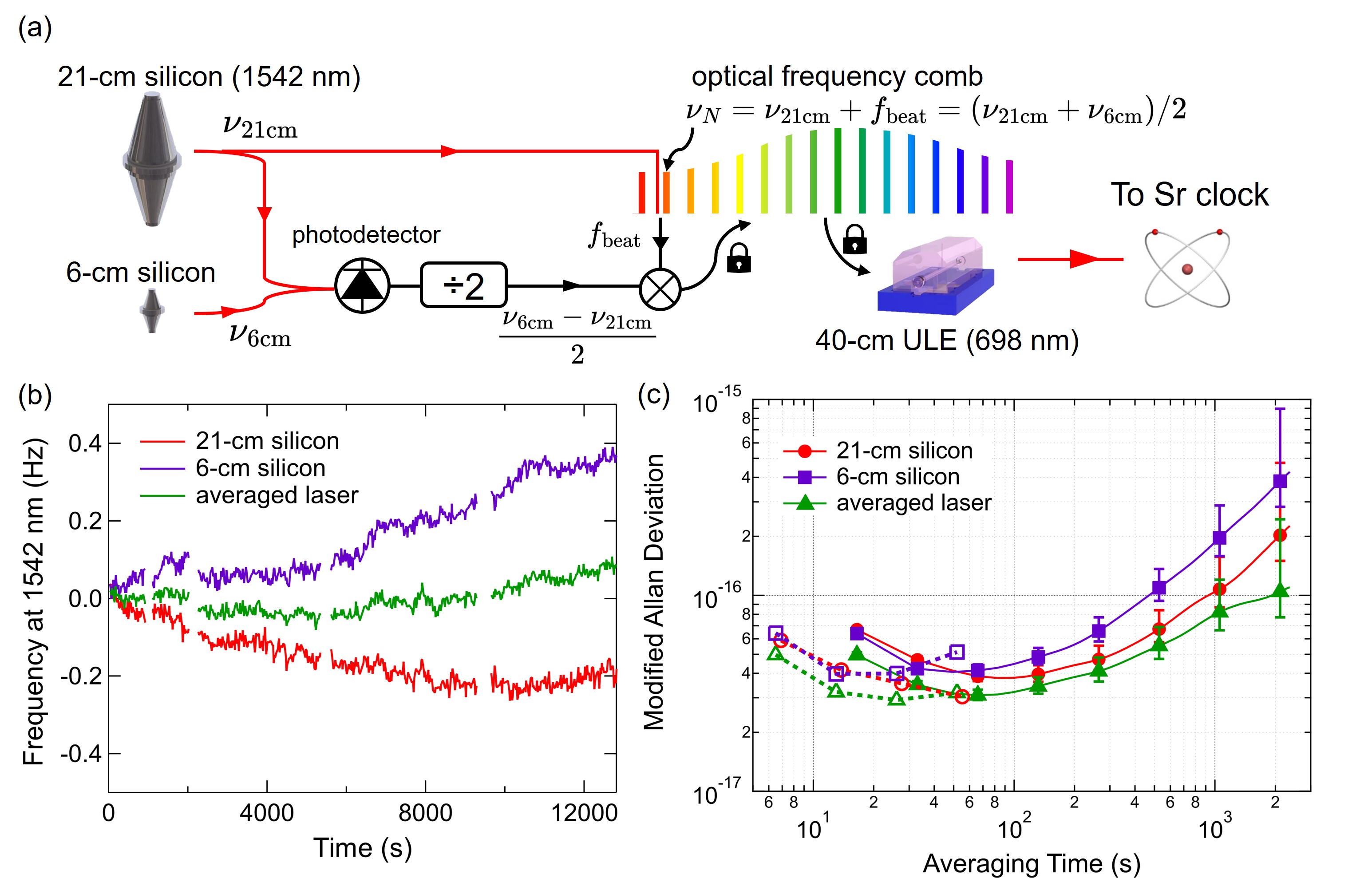}
\caption{\label{fig:averaging}Optical frequency averaging of light using two silicon cavities. (a) Schematic diagram of the setup. By phase locking $f_\text{beat}$, the beat note between the comb and the 21-cm silicon cavity, to a separately generated $(\nu_\text{6cm}-\nu_\text{21cm})/2$, the comb is stabilized to the average frequency of the two silicon cavities. A 698~nm laser locked to the 40-cm ULE cavity is further stabilized to the comb and interrogates Sr atoms. (b) Frequency record of the lasers locked to the 21-cm cavity, 6-cm cavity, and the averaged laser, measured with the Sr clock. (c) Modified Allan deviation of the three lasers computed from the data in (b) (solid lines). The relatively high instability at short averaging times is from Dick noise, which is verified by operating the clock with a shorter dead time, hence less Dick noise (dotted lines).}
\end{figure*}

Fig.~\ref{fig:averaging}(a) shows how optical frequency averaging is implemented. A 1542-nm laser stabilized to the 21-cm cavity is heterodyned with a self-referenced Er:fiber frequency comb such that the nearest comb mode frequency can be written as $\nu_N=\nu_\text{21cm}+f_\text{beat}$, where $\nu_N$ is the optical frequency of the $N^\text{th}$ comb mode, $\nu_\text{21cm}$ is the optical frequency of the laser locked to the 21-cm silicon cavity, and $f_\text{beat}$ is the RF beat note between the two optical frequencies. For normal operation, $f_\text{beat}$ is phase locked to a synthesizer by feeding back to the comb, thereby transferring the stability of the 21-cm silicon cavity to the comb. A 698-nm laser pre-stabilized to the 40-cm ULE cavity is then phase locked to the comb and interrogates Sr atoms. For optical frequency averaging, a simple modification is made to the phase locking scheme. Instead of locking $f_\text{beat}$ to a synthesizer, it is locked to $(\nu_\text{6cm}-\nu_\text{21cm})/2$ generated with an RF frequency divider, where $\nu_\text{6cm}$ is the optical frequency of the 6-cm silicon cavity. With this modification, the comb mode frequency becomes $\nu_N=(\nu_\text{21cm}+\nu_\text{6cm})/2$, i.e., the average frequency of the two cavities. The comb can also be locked to the 6-cm cavity by bypassing the divide-by-two operation.

The frequencies of the lasers locked to the two silicon cavities and their average frequency, measured with the Sr lattice clock are shown in Fig.~\ref{fig:averaging}(b). The laser frequencies are measured with three independent, interleaved feedback loops that steer the laser frequencies to the Sr clock transition. The interleaving is achieved by switching the laser that interrogates the Sr atoms for consecutive cycles of the Sr clock. The action of the averaging operation is clearly seen in the cancellation of the opposite-sign drifts of the two cavities. The solid lines in Fig.~\ref{fig:averaging}(c) show the modified Allan deviation corresponding to the data in Fig.~\ref{fig:averaging}(b). The averaged laser is more stable than the two similar individual lasers by approximately $\sqrt{2}$, consistent with expectation. The measurement at short averaging times is limited by Dick noise due to the long dead time inherent to interleaved measurements \cite{dick1989local}. To verify this, we reduce the dead time by measuring the three lasers independently against Sr atoms without interleaving, shown in dashed lines. As expected, the Dick noise contribution is reduced at shorter averaging times.

A more stable local oscillator corresponds to an improved clock stability for a Dick noise-limited clock such as ours, evidenced by the reduced Dick noise of the averaged laser in Fig.~\ref{fig:averaging}(c). Because of the improved clock stability, evaluation of systematic clock uncertainties can be performed twice faster with a $\sqrt{2}$ times lower instability of the local oscillator. Furthermore, the longer laser coherence time allows a longer clock interrogation time, reducing the instability due to quantum projection noise \cite{marshall2025high, kim2023improved}.

\textbf{\emph{Drift of silicon cavities}.} The drift rate of cryogenic silicon cavities is orders of magnitude lower than that of ULE cavities. While typical ULE cavities drift a few kilohertz per day, cryogenic silicon cavities drift only a few hertz per day. It is currently unknown where this small amount of drift comes from. Unlike ULE cavities whose spacers are made of amorphous glass material that can relax over time, silicon cavities are made of crystalline material and therefore should not drift at all in principle. Silicon cavities with crystalline coatings are especially interesting because even the mirror coatings are crystalline, making the entire cavity crystalline. To shed light on the origin of the drift, we report the long-term drift rates of four silicon cavities currently operating at PTB and JILA, accumulated over more than 10 years. The naming convention of the four cavities and their relevant features, as well as previous publications, are summarized in Table~\ref{tab:cavities}.
\begin{table}
\caption{\label{tab:cavities}%
Four cryogenic silicon cavities at JILA and PTB whose drift rates are reported in this work.}
\begin{ruledtabular}
\begin{tabular}{ccccc}
 Name & Length & Temp. & Mirror material & Refs. \\
\hline
Si2 & 21~cm & 124 K & $\text{Si}\text{O}_2/\text{Ta}_2\text{O}_5$ & \cite{matei2017} \Tstrut\\
Si3 & 21~cm & 124 K & $\text{Si}\text{O}_2/\text{Ta}_2\text{O}_5$ & \cite{matei2017, oelker2019demonstration} \\
Si5 & 21~cm & 124 K &  $\text{GaAs}/\text{Al}_{0.92}\text{Ga}_{0.08}\text{As}$ & \cite{kedar2023frequency, yu2023excess} \\
Si6 & 6~cm & 17 K &$\text{GaAs}/\text{Al}_{0.92}\text{Ga}_{0.08}\text{As}$ & \cite{kedar2023frequency, yu2023excess} \\
\end{tabular}
\end{ruledtabular}
\end{table}

\begin{figure}
\includegraphics[width=\linewidth]{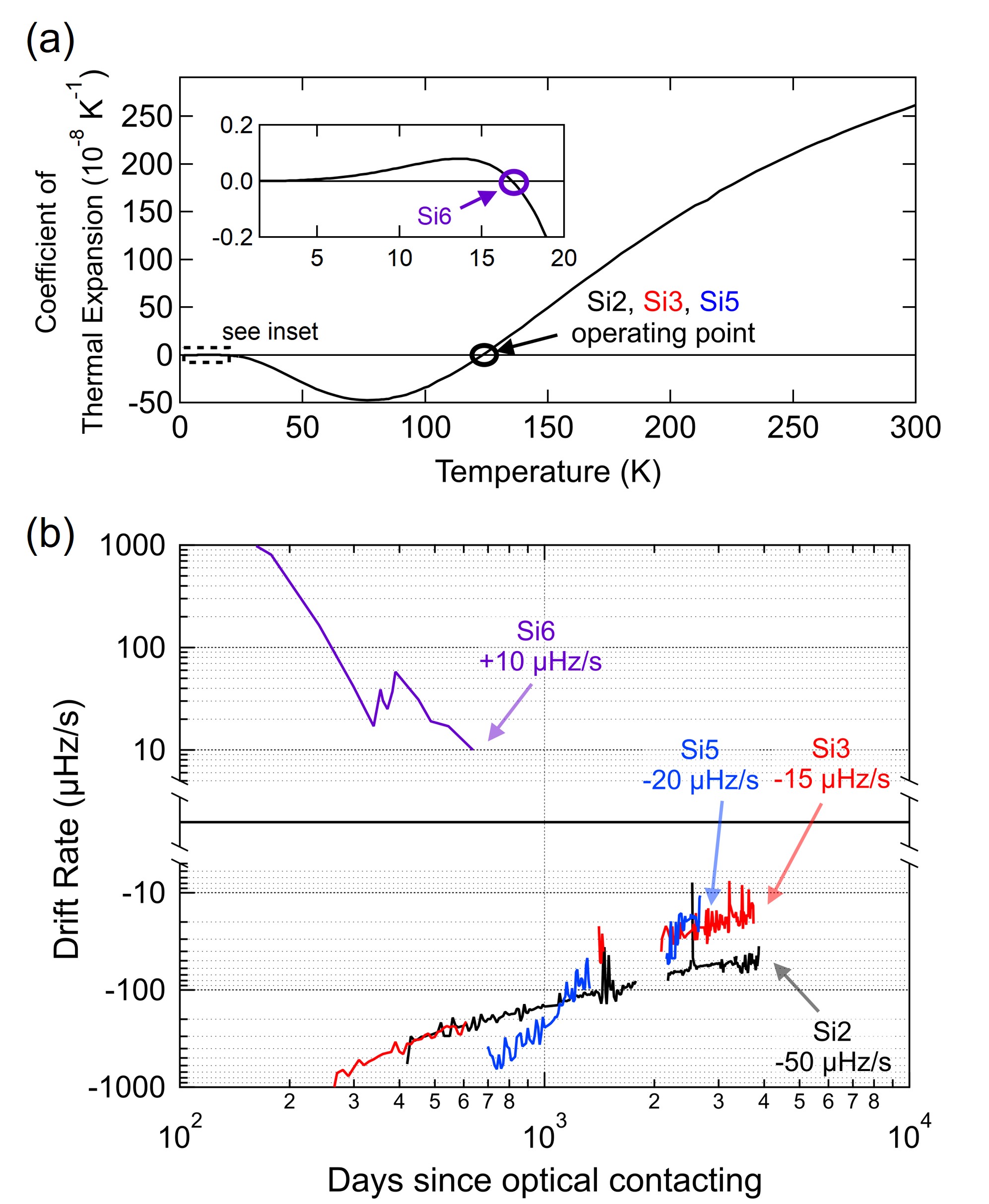}
\caption{\label{fig:drift}Coefficient of thermal expansion of silicon and long-term drift rates of the four cryogenic silicon cavities in Table~\ref{tab:cavities}. (a) Coefficient of thermal expansion of silicon. The data between 1.5~K and 24~K are from \cite{wiens2014silicon} and between 24~K and 300~K are from \cite{lyon1977linear}. (b) Drift rates of the four silicon cavities (Black: Si2; Red: Si3; Blue: Si5; Purple: Si6). All four cavities exhibit very low drift rates in the tens of microhertz per second range.}
\end{figure}

Fig.~\ref{fig:drift}(a) shows the coefficient of thermal expansion of silicon over a range of temperatures \cite{wiens2014silicon, lyon1977linear}. To suppress cavity length changes from temperature fluctuations, the silicon cavities operate at zero crossings of the coefficient of thermal expansion. Three of the four cavities reported here (Si2, Si3, Si5) operate at the 124~K zero crossing. Si6, on the other hand, operates at the 17~K zero crossing (Fig.~\ref{fig:drift}(a) inset), where the much gentler slope of the coefficient of thermal expansion significantly relaxes the requirements on temperature stability.

The long term drift rates of the silicon cavities are shown in Fig.~\ref{fig:drift}(b). The origin of the horizontal axis is chosen to be the day each cavity was assembled by optical contact bonding the mirrors to the spacer, as a representative for the ``age'' of the cavity. The absolute frequencies of the cavities are tracked with hydrogen masers or optical clocks. For Si2, Si3, and Si5, the drift rates are calculated by first computing a 10-day binned average of the cavity frequency and then taking the derivative. For Si6, the system was still being optimized for a large portion of the dates plotted here such that 10-day averaging was not appropriate. The Si6 drift rate is therefore calculated over durations ranging from 1 day to 24 days. After several years, all of the silicon cavities settle to drift rates in the tens of microhertz per second range. Some distinguishing features can be observed in the drift rate of the 6-cm cavity Si6, although the record to date spans less than 2 years. First, the sign of the drift rate is opposite from that of all the other silicon cavities. The effective cavity length of Si6 is getting shorter over time, similar to the drift behavior of glass and ceramic cavities \cite{alvarez2019optical, legero2010tuning, hall1993frequency, dube2009narrow, zhang2024cryogenic, ito2017stable}. Second, the drift rate of Si6 settles to a low value much faster than the other cavities. All three other cavities took several years until the drift rate reached tens of microhertz per second level whereas Si6 took less than two years before reaching a similar drift rate.

To put these drift rates into context, we note that the absolute frequency of Si2, the cavity whose frequency has been tracked the longest, has drifted only $-44~\text{kHz}$ in the past 10 years. Converted to cavity length change, this corresponds to an overall lengthening of the 21-cm cavity by 48~pm, which is approximately $1/11$ of the silicon lattice constant. The recent drift rates of $-50~\mu\text{Hz/s}$ ($-2.6 \times 10^{-19}~\text{s}^{-1}$ in fractional units), $-15~\mu\text{Hz/s}$ ($-7.7 \times 10^{-20}~\text{s}^{-1}$), $-20~\mu\text{Hz/s}$ ($-1.0 \times 10^{-19}~\text{s}^{-1}$), and 10~$\mu\text{Hz/s}$ ($5.2 \times 10^{-20}~\text{s}^{-1}$) of Si2, Si3, Si5, and Si6, respectively, are nearly competitive with typical fractional drift rates of active hydrogen masers in the $10^{-21}-10^{-20}~\text{s}^{-1}$ range \cite{parker2010medium, parker1999hydrogen, griffin2021drift}. Further stability improvements in the range of $10^4$ to $10^6$~s will open up the possibility of an all-optical timescale using optical cavities and optical clocks \cite{milner2019demonstration, mcgrew2019towards, grebing2016realization, yao2019optical}.

As mentioned previously, the mechanism that causes the drift in cryogenic silicon cavities, let alone its sign and settling behavior, is currently unknown. Possible mechanisms include slow relaxation of the stress induced by cavity mounting structure, optically contacted surfaces \cite{berthold1976dimensional, hall1993frequency}, or mirror coatings \cite{kreider2025quantification}. Further investigations and new cryogenic silicon cavities at different operating temperatures might provide more insight into the origin of this drift behavior \cite{barbarat2024towards, wiens2016resonator}.

\textbf{\emph{Conclusion}.} The fractional frequency stability of cavity-stabilized lasers now reach an impressive $10^{-17}$ level, yet still limits the stability of optical clocks. An important step towards next-generation optical cavities is the development of low mechanical loss semiconductor crystalline mirrors. In this work, we demonstrate a record fractional frequency stability of $2.5 \times 10^{-17}$ on a 6-cm cryogenic silicon cavity with crystalline AlGaAs mirrors. We extract an upper bound on the mechanical loss factor of $2.3 \times 10^{-5}$ at 17~K, confirming expectations on the superior mechanical properties of crystalline AlGaAs coatings over dielectric coatings. We also demonstrate optical frequency averaging of two silicon cavities, resulting in an optical frequency that is more stable than its constituents both at short and long averaging times. In addition, the long-term drift rates of four silicon cavities over several years are reported. By combining the key properties already realized in this and other silicon cavities, namely, 21~cm length, 17~K operation, large radius of curvature mirrors, and crystalline coatings, we anticipate a cryogenic silicon cavity with low-$10^{-18}$ performance to be practically feasible.

\begin{acknowledgments}
We thank Y. Yang, E. Y. Song, and S. Agrawal for providing insightful comments on the manuscript. We acknowledge contributions and discussions from G. D. Cole, G.-W. Truong, and W. Warfield. Funding for this work is provided by NSF QLCI OMA-2016244, V. Bush Fellowship, NSF PHY-2317149, and NIST. PTB acknowledges support by the Deutsche Forschungsgemeinschaft (DFG, German Research Foundation) under Germany’s Excellence Strategy–EX-2123 QuantumFrontiers (Project No. 390837967) and by the Max Planck-RIKEN-PTB Center for Time, Constants and Fundamental Symmetries. 
\end{acknowledgments}

\appendix
\section{\label{Append:thermal noise}Thermal noise calculation}
For cryogenic silicon cavities, the spacer and substrates contribute a negligible amount of noise to the total thermal noise, leaving the coating Brownian noise as the only dominant term. We use the formula from \cite{harry2006thermal}, also used in \cite{cole2013tenfold, robinson2021thermal}, to calculate the coating Brownian noise:
\begin{equation}
\label{eqn:TN}
\begin{split}
S_y(f) = &\frac{4k_BTd_\text{coat}\phi_\text{coat}}{\pi^2w^2Y_\text{sub}L^2f}\left[\frac{Y_\text{coat}}{Y_\text{sub}}\frac{(1+\sigma_\text{sub})^2(1-2\sigma_\text{sub})^2}{1-\sigma_\text{coat}^2} \right. \\
& \left. + \frac{Y_\text{sub}}{Y_\text{coat}}\frac{(1+\sigma_\text{coat})^2(1-2\sigma_\text{coat})}{1-\sigma_\text{coat}^2}\right]~,
\end{split}
\end{equation}
where the meaning of each symbol and its value are summarized in Table~\ref{tab:constants}.
\begin{table}
\caption{\label{tab:constants}
Symbols used in Eq.~\eqref{eqn:TN}}
\renewcommand{\arraystretch}{1.3}
\begin{ruledtabular}
\begin{tabular}{ccc}
 Symbol & Description & Value \\
\hline
$S_y(f)$ & \begin{tabular}{@{}c@{}}fractional frequency noise \\ power spectral density ($\text{Hz}^{-1}$)\end{tabular} &  \Tstrut\\
$f$ & Fourier frequency (Hz) &   \\
$k_B$ & Boltzmann constant & $1.38 \times 10^{-23} ~\text{J}\,\text{K}^{-1}$ \\
$T$ & temperature & 17~K \\
$d_\text{coat}$ & coating thickness & $12.0~\mu\text{m}$\\
$w$ & beam radius at the mirror & 293~$\mu$m \\
$L$ & cavity length & 6.02~cm \\[0.1cm]
$Y_\text{coat}$ & Young's modulus of coating\footnote{\cite{robinson2021thermal} for dielectric, \cite{cole2013tenfold} for crystalline}& \footnotesize \begin{tabular}{@{}c@{}} 100~GPa (crystalline) \\ 91~GPa (dielectric) \end{tabular} \\[0.3cm]
$Y_\text{sub}$ & Young's modulus of substrate\footnote{along the $\langle111\rangle$ direction of silicon} & 188~GPa \\[0.1cm]
$\sigma_\text{coat}$ & Poisson's ratio of coating\footnote{\cite{robinson2021thermal} for dielectric, \cite{cole2013tenfold} for crystalline} & \begin{tabular}{@{}c@{}} 0.32 (crystalline) \\ 0.2 (dielectric) \end{tabular} \\[0.3cm]
$\sigma_\text{sub}$ & Poisson's ratio of substrate & 0.26\\
$\phi_\text{coat}$ & mechanical loss factor of coating & see Table~\ref{tab:loss}.
\end{tabular}
\end{ruledtabular}
\end{table}
The mechanical loss factors for dielectric and crystalline coatings are listed in Table~\ref{tab:loss}.
\begin{table}
\caption{\label{tab:loss}}
Mechanical loss factors of dielectric $\text{Si}\text{O}_2/\text{Ta}_2\text{O}_5$ and crystalline $\text{GaAs}/\text{Al}_{0.92}\text{Ga}_{0.08}\text{As}$ coatings
\renewcommand{\arraystretch}{1.5}
\begin{ruledtabular}
\begin{tabular}{ccc}
Temperature & Dielectric & Crystalline \\
\hline
300~K & $4 \times 10^{-4}$ \cite{numata2004thermal} & $2.5 \times 10^{-5}$ \cite{penn2019mechanical} \Tstrut\\[0.2cm]
124~K & $2.4 \times 10^{-4}$ \cite{robinson2021thermal} & $2.5 \times 10^{-5}$ \cite{yu2023excess} \\[0.2cm]
$<17~\text{K}$ & $3.2 \times 10^{-4}$ \cite{robinson2021thermal} &
\renewcommand{\arraystretch}{1.2}
\begin{tabular}{@{}c@{}}
$5 \times 10^{-5}$ \cite{cole2008monocrystalline} \\
$6 \times 10^{-6}$ \cite{cole2012cavity} \\
$4 \times 10^{-5}$ \cite{pag25} \\
$<2.3 \times 10^{-5}$ [this work] \end{tabular} \\
\end{tabular}
\end{ruledtabular}
\end{table}
The $1/f$ dependence of $S_y(f) = h_{-1}f^{-1}$ leads to a constant modified Allan deviation of $\text{mod}\,\sigma_y \approx \sqrt{0.936\,h_{-1}}$ \cite{stein1985frequency}, shown in Fig.~\ref{fig:mdev} as the thermal noise limit.

\bibliography{main}

\end{document}